\documentclass[11pt]{article}

\usepackage{hyperref}
\pdfoutput=1
\usepackage[applemac]{inputenc}

\begin{document}
\title{Inside a kettle}
\author{S. Wildeman, H. Lhuissier, C. Sun \& D. Lohse\\ \\\vspace{6pt} POF - University of Twente \\ The Netherlands}

\maketitle

%% The abstract (in this file, and that submitted as text to arXiv) should include the exact phrase %% "fluid dynamics video" or "fluid dynamics videos"
\begin{abstract}
	This fluid dynamics video images the different heat transport mechanisms at play when a liquid confined in a vertical Hele-Shaw cell is heated from below. The two-dimensional time resolved temperature field inside the cell is measured by a quantitative Schlieren technique which is detailed in the video.
\end{abstract}

% main text
%\section{Introduction}
	When a liquid in a container is heated from below and cooled from the top a heat flux settles through the liquid. As the temperature of the bottom plate increases, the heat transport is successively dominated by different mechanisms: conduction, convection and boiling. To study this, experiments with a Hele-Shaw cell were undertaken. Instantaneous non-intrusive measurements of the two-dimensional temperature field are performed in the cell. The video presents these measurements in combination with Schlieren visualizations of the flow to illustrate the different heat transport mechanisms and the transitions between them.
	
	The cell is made of two thin vertical glass slides ($50\,\rm{mm}$ in width and $25\,\rm{mm}$ in height) separated by a $1\,\rm{mm}$ gap. The bottom of the cell is made of a thin heat-conducting plate. The top is open. The movie begins when a cell filled with ethanol at ambient temperature is put on top of a hot copper block. In order to visualize the temperature evolution inside the cell a screen patterned with a regular cross-ruling is put behind the cell. As the liquid in the cell warms up, the cross-ruling -- as seen through the cell -- looks distorted. The temperature differences in the cell plane give rise to index of refraction gradients. This bends the light rays, resulting in a distorted image of the cross-ruling. Because the index of refraction-temperature relationship and the distance to the screen are known, measurements of the distortion can be used to calculate the temperature gradients in the cell.
	
	By integrating those gradients, this technique provides the temperature field inside the cell with a time resolution that is only limited by the camera frame rate (here 500 frames per second). This is used in the video to illustrate the different heat transport mechanisms at play as the liquid inside the cell is heated from below to progressively higher temperatures. One successively observes the temperature fields for:
\begin{itemize}	
	\item two-dimensional convection motions with light, hot plumes rising between falling heavy, cold ones,
	\item nucleate boiling of vapor bubbles on the bottom plate as the ethanol boiling temperature is exceeded,
	\item and the `boiling crisis', i.e. the transition to the film boiling regime where the bottom plate is covered by a vapor film which periodically destabilizes by a Rayleigh-Taylor instability.
\end{itemize}

\end{document}